\documentclass[apj]{emulateapj}

\usepackage{graphicx,times}
\usepackage{subfigure}
\newcommand{\be}{\begin{equation}}
\usepackage{threeparttable}
\usepackage{booktabs}
\newcommand{\ee}{\end{equation}}
\newcommand{\bea}{\begin{eqnarray}}
\newcommand{\eea}{\end{eqnarray}}

\usepackage{amsmath}
\usepackage{cases}
\usepackage{longtable}
\usepackage{hyperref}
\usepackage{epstopdf}
\usepackage{amsmath,bm}
\usepackage{amssymb}
\usepackage{natbib}
\usepackage{morefloats}
\usepackage{multirow}
\usepackage{array}
\usepackage{verbatim}

\begin{document}

\title{On the origin of the scatter broadening of fast radio burst pulses and astrophysical implications}

\author{Siyao Xu\altaffilmark{1} and Bing Zhang \altaffilmark{1,2,3}}

\altaffiltext{1}{Department of Astronomy, School of Physics, Peking University, Beijing 100871, China; syxu@pku.edu.cn}
\altaffiltext{2}{Kavli Institute for Astronomy and Astrophysics, Peking University, Beijing 100871, China}
\altaffiltext{3}{Department of Physics and Astronomy, University of Nevada Las Vegas, NV 89154, USA; zhang@physics.unlv.edu}

\begin{abstract}

Fast radio bursts (FRBs) have been identified as extragalactic sources which can make
a probe of turbulence in the 
intergalactic medium (IGM) and their host galaxies. 
To account for the observed millisecond pulses caused by scatter broadening, we examine a variety of possible models of electron density fluctuations in both the 
IGM and the host galaxy medium. 
We find that a short-wave-dominated
power-law spectrum of density, which may arise in highly supersonic turbulence with pronounced local dense structures of shock-compressed gas
in the host interstellar medium (ISM), can produce the required density enhancements at sufficiently small scales to interpret the scattering timescale of FRBs.
It implies that an FRB residing in a galaxy with efficient star formation in action tends to have a broadened pulse.
The scaling of the scattering time with dispersion measure (DM) in the host galaxy varies in different turbulence and scattering regimes.
The host galaxy can be the major origin of scatter broadening, but contribute to a small fraction of the total DM.
We also find that
the sheet-like structure of density in the host ISM associated with folded magnetic fields in a viscosity-dominated regime of 
magnetohydrodynamic (MHD) turbulence cannot give rise to strong scattering.
Furthermore, valuable insights into the IGM turbulence concerning the detailed spatial structure of density and magnetic field 
can be gained from the observed scattering timescale of FRBs. 
Our results are in favor of the suppression of micro-plasma instabilities and the validity of collisional-MHD 
description of turbulence properties in the collisionless IGM.
%This finding provides information on the detailed spatial structure of extragalactic magnetic fields.

\end{abstract}


\section{Introduction}

A population of bright millisecond radio transients known as fast radio bursts (FRBs) 
have been discovered and attracted increasing attention in recent years 
(e.g.  \citealt{Lor07, Tho13, Mas15, Kea16, Pet16, Spi16}).
The large dispersion measure (DM) values and high Galactic latitudes of these events 
provide strong observational evidence of their extragalactic origin
\citep[e.g.][]{Kat16}.
%Direct redshift measurement further enables a reliable determination of the cosmological distance
%\citep{Kea16}.

As one of the important observational parameters of FRBs, the pulse broadening time scale (i.e. pulse width with the intrinsic timescale subtracted) is a result of the 
multi-path scattering during the propagation of radio waves through a turbulent medium. 
The Galactic contribution in pulse broadening can be easily eliminated since the Galactic pulsars at high latitudes visually possess 
orders of magnitude smaller broadening timescales than FRBs
\citep{Bha04, Kri15, Kat16, Cor16}
\footnote{For the low-latitude FRBs, i.e.
FRB 010621
\citep{Kea12},
FRB 150418
\citep{Kea16}, 
FRB 121102
\citep{Spi14},
only upper limits on the broadening time are available. Even for these FRBs,
the NE2001 model of Galactic scattering 
\citep{Cor02}
predicts that the Galactic contribution to the scattering timescale is below the threshold of detectability.}.
Non-Galactic contributions may arise from the IGM and the host galaxy medium. 
The empirical relation between the scattering measure and DM in the IGM estimated by 
\citet{Mac13}
demonstrates that the scattering per unit DM in the IGM is orders of magnitude smaller than that in Galactic ISM. 
The possibility of prominent intergalactic scattering was disputed by 
\citet{LG14}
because of the incompatibility between the excessive heating of Kolmogorov turbulence with a small outer scale
and inefficient cooling of the IGM.
The IGM was also disfavored as the location of scattering by 
\citet{Ka16, Kat16}
based on the non-monotonic dependence of pulse widths on intergalactic dispersion.
\footnote{One caveat of this argument is that the empirical scattering measure-DM relationship in Milky Way has a large dispersion \citep{Jo98, Bha04}. However, after correcting for such a scatter, \cite{Ca16} still could not interpret the FRB scattering data. }
Apart from the above arguments, 
according to the catalog of known FRBs provided by 
\citet{Pet16}, 
some FRBs have scattering time longer than 1 ms (at 1 GHz), 
while the others have unresolved scattering tails, for which the upper limit is set by the time resolution, but the actual scattering can be 
much weaker ($\ll 1$ ms).
Intuitively, the observational facts that some FRBs have greater DMs but narrower pulses
and that both resolved and unresolved pulses exist
imply that the scatter broadening is not a common feature originating from the IGM that every FRB pulse traverses through, 
but more likely attributed to the diverse environments local to FRBs.
That is, the host galaxy is the most promising candidate for interpreting the strong scattering events
(see \citealt{Yao16} for a different point of view).
However, the host contribution depends on the progenitor location and line-of-sight (LOS) inclination. It is expected to be negligibly small  
for sightlines passing through a host galaxy's outskirts, similar to the case of our Galaxy at high Galactic latitudes. 
For this reason, it was suggested that the pulse broadening is produced by the highly turbulent and dense medium in the immediate vicinity 
of the FRB
\citep{Kat16}.
But since the scattering material is in strong association with the burst, the resulting pulse width is likely entirely intrinsic, and 
the scenario is restricted to specific FRB progenitor models involving young stellar populations
\citep{Mas15, Spi16}.

A proper interpretation of the temporal broadening of FRBs entails comprehensive modeling of the electron density fluctuations and related turbulence 
properties in both the ISM and IGM. 
A Kolmogorov spectrum of both velocity and magnetic fluctuations was predicted by the 
\citet{GS95}
theory for Alfv\'{e}nic turbulence and later confirmed by MHD simulations 
\citep{CV00, MG01, CL03, BL09}.
The observationally measured electron density power spectrum in the diffuse ionized ISM 
is also consistent with a Kolmogorov-like power law over a wide range of scales spanning over 10 decades,
known as ``the big power law in the sky"
\citep{Armstrong95, CL10}.
In earlier studies on the scatter broadening of FRB pulses
(e.g., \citealt{Mac13, LG14, Cor16}),
the Kolmogorov model of turbulence has been commonly adopted. 
However, the spectral form of density fluctuations can be affected by the magnetization and compressibility of the local turbulent medium. 
The density fluctuations do not track the Kolmogorov velocity spectrum, 
but exhibit a steeper spectrum in a strongly magnetized subsonic turbulence, and a shallower one in supersonic turbulence
\citep{BLC05, KL07}.
As a general result of both compressible MHD simulations and hydrodynamic simulations, 
supersonic turbulence effectively generates a complex system of shocks which correspond to 
regions of converging flows and concentration of mass
\citep{PaJ01, PJ04, Kim05, Kri06}.
\citet{Kim05}
explicitly showed that the density power spectrum becomes shallower with increasing sonic Mach number $M_s$.
Notice that $M_s$ varies in different ISM phases.
The warm ionized medium (WIM) has $M_s$ of order unity 
\citep{Haf99, Kim05, Hi08}
and hence Kolmogorov density distribution
\citep{Armstrong95, CL10}, 
while in other colder and denser phases in the inner Galaxy with a higher compressibility (i.e., larger $M_s$,
\citealt{Lar81,Hei03}),
a shallower density spectrum is naturally expected
\citep{Kim05, Bur15}.
Significant deviation from the Kolmogorov law and flattening of the density spectrum are indicated from e.g., spectroscopic observations 
\citep{Las06,Laz09rev,HF12},
scattering measurements of the Galactic pulsars with high DMs
(\citealt{Loh01, Loh04, Bha04, Lew13, Lew15, Kri15}; Xu \& Zhang 2016, in preparation), 
and rotation measure fluctuations at low Galactic latitudes 
\citep{Hav04, Hav08, XuZ16}.
Accompanying the shallowness of the spectral slope of density fluctuations, 
substantial discontinuous structures in density emerge at small scales due to supersonic compressions.
The corollary is to significantly strengthen the scatter broadening effect.
Besides the spectral slope, 
the distinct properties of turbulence in different ISM phases are also manifested in the volume filling factor of density structures. 
The volume filling factor of cold and dense phases, such as the cold neutral medium and molecular clouds, 
is smaller than that of the WIM by order(s) of magnitude
\citep{Tie,HS13}.
The small-scale overdense structures embedded in these phases produced by the supersonic turbulence are supposed to have a further smaller filling factor.

In view of the theoretical arguments and observational facts, we consider the spectrum of density fluctuations 
with a much shallower slope than the Kolmogorov one
as a physically motivated possibility of inducing enhanced scattering. 
Moreover, we also take into account the 
microscale density fluctuations associated with the microphysical properties of turbulence, 
which include
the density perturbations caused by the mirror instability in the collisionless regime of MHD turbulence
\citep{Ha80},
and the sheet-like configuration of density generated by the magnetic folds in the viscosity-damped regime of MHD turbulence 
\citep{GS06, Laz07}.
We will examine whether they can serve as an alternative source of strong scattering.

In this work, to identify the separate roles of the IGM and host galaxy in temporal smearing and 
probe the environmental conditions of FRBs, 
we examine the scattering effect of different models of electron density fluctuations pertaining to distinct turbulence regimes, 
including a detailed analysis on both the Kolmogorov and shallower density power spectra, 
and an exploratory investigation on other not well-determined but potentially important models of density structures. 
On the other hand, with the radio signals traveling across cosmological distances,  
%not only does the broadened pulse of FRBs appeal to a suitable description of density fluctuations, 
the investigation on the scatter broadening of FRBs offers a promising avenue for probing the IGM turbulence,
which remains a highly controversial and elusive subject concerning whether a collisional-MHD treatment is still valid 
for the dynamics of the weakly collisional IGM 
\citep{Rei14}
or the large-scale dynamics is dramatically affected by the microscale instabilities
\citep{Sch08}.

This paper is organized into four sections. In Section 2, we focus on the 
power-law model of electron density fluctuations and the effect of shallowness of spectral slope on temporal broadening. 
In Section 3, we generalize the analysis and evaluate the scattering strength of other alternative models of electron density fluctuations 
on the basis of the observed scattering timescale. 
Implications and conclusions are presented in Section 4.

\section{Electron density fluctuations arising from a turbulent cascade}
\label{eq: secpw}

\subsection{Temporal broadening}

A power-law spectrum of the plasma density irregularities is commonly applied in studies on radio wave propagation
\citep{Lee76, Rick77, Ric90}, 
which is also reinforced by growing observational evidence of interstellar density fluctuations 
\citep{Armstrong95, CL10}.
We assume that the scattering effect is introduced by electron density fluctuations that arise from a turbulent cascade and the relevant spectrum takes 
the form 
\citep{Rick77, Col87}
\begin{equation}\label{eq: orisf}
    P(k) = C_N^2 k^{-\beta} e^{-(kl_0)^2}, ~~ k> L^{-1}, 
\end{equation}
which is cast as a power-law spectrum in the inertial range of turbulence, 
\begin{equation}
    P(k) = C_N^2 k^{-\beta}, ~~ L^{-1}<k<l_0^{-1}, 
\end{equation}
with $L$ and $l_0$ as the outer and inner scales. 
%corresponding to the injection and dissipation scales of turbulent energy. 
The spectral index $\beta$ is suggested to be within the range $2< \beta < 4$ on observational grounds 
(e.g., \citealt{Lee75, Rick77, Ro86}). 
Intuitive insight to the properties of turbulence can be gained from the value of $\beta$. 
At the critical index $\beta =3$, density fluctuations, which scale as $\delta n_e \propto k^{(3-\beta)/2}$, are scale-independent.
That is, the density fluctuations with the same amplitude exist at all scales. 
Notice that $\delta n_e$ represents the root-mean-square (rms) amplitude of density fluctuations.
Following the power-law statistics studied in e.g., 
\citet{LP00,LP04,EL05,LP06},
we consider the density spectrum in both the long-wave-dominated regime with $\beta>3$ and the  
short-wave-dominated regime with $\beta<3$.
The density field in the former case is dominated by large-scale fluctuations, but in the latter case is localized in small-scale structures.

Both long- and short-wave-dominated density spectra are a confirmed reality in compressible MHD turbulence
\citep{BLC05, KL07}.
In the WIM phase of Galactic ISM which corresponds to the transonic turbulence,
the power-law spectrum of electron density fluctuations has been convincingly demonstrated 
to have a unique slope consistent with the Kolmogorov spectrum ($\beta = 11/3$) on scales spanning from $10^6$ to $10^{17}$ m 
\citep{Armstrong95,CL10}.
On the other hand, in colder and denser phases of the ISM in the Galactic plane, the turbulence becomes highly supersonic and shocks are 
inevitable, which produce large density contrasts and a short-wave-dominated density spectrum
\citep{Kim05}.
The density spectra with $\beta<3$ have been extracted from ample observations by using different tracers and techniques 
(e.g., \citealt{Stu98,Des00,Sw06,XuZ16}; also see table 5 in the review by \citealt{Laz09rev} and figure 10 in the review by \citealt{HF12}).
In addition, the Kolmogorov density spectrum also fails to reconcile with the observationally measured
scaling relation between scatter-broadening time and frequency for highly dispersed pulsars 
(e.g., \citealt{Loh01, Loh04, Bha04, Lew13}).
In view of the diversity of ISM phases and properties of the associated turbulence, 
it is necessary to perform a general analysis incorporating both long- and short-wave-dominated spectra of density fluctuations.

The normalization of the power spectrum depends on the steepness of the slope. From the density variance 
\begin{equation}
    \langle (\delta n_e)^2 \rangle = \int P(k) d^3 \bm{k}
\end{equation}
and assuming $L \gg l_0$, we find 
\begin{subnumcases}
    {C_N^2 \sim \label{eq: cnsss}}
     \frac{\beta-3} {2(2\pi)^{4-\beta}}(\delta n_e)^2 L^{3-\beta},   ~~ \beta > 3, \label{eq: cnstep} \\
     \frac{3-\beta}{2(2\pi)^{4-\beta}}(\delta n_e)^2 l_0^{3-\beta}, ~~~ \beta < 3.\label{eq: cnshal}
\end{subnumcases}
It shows that the turbulent power characterized by density perturbation $\delta n_e$
concentrates at $L$ for $\beta>3$ and $l_0$ for $\beta<3$. 
Thus $\delta n_e$ is the density perturbation at the correlation scale of turbulence, which is $L$ for a 
long-wave-dominated spectrum and $l_0$ for a short-wave-dominated spectrum.

As the radio waves propagate through a turbulent plasma, multi-path scattering causes temporal broadening of a transient pulse
(e.g., \citealt{Ric90,CR98}).
On a straight-line path of length $D$ through the scattering medium, the integrated phase structure function is defined as the 
mean square phase difference between a pair of LOSs with a separation $r$ on the plane transverse to the propagation direction,
$D_\Phi = \left\langle(\Delta \Phi)^2\right\rangle$.
Given the spectral form of Eq. \eqref{eq: orisf} with $2<\beta<4$, and under the condition $r\ll L\ll D$, $D_\Phi$ has expressions 
\citep{Col87,Ric90}
\begin{subnumcases}
    {D_\Phi \sim \label{eq: sfspec}}
     \pi r_e^2 \lambda^2 \text{SM} l_0^{\beta-4} r^2,  ~~ r<l_0, \\
     \pi r_e^2 \lambda^2 \text{SM} r^{\beta-2}, ~~~~~ r>l_0, \label{eq: sflp}
\end{subnumcases}
where $r_e$ is the classical electron radius and $\lambda$ is the wavelength.
The scattering measure SM is the integral of $C_N^2$ along the LOS path through the scattering region, and characterizes the 
scattering strength. 
Here we consider a statistically uniform turbulent medium, with the turbulence properties independent of the path length.
Thus the SM is simplified as
\begin{equation}\label{eq: simsm}
     \text{SM}  \sim  C_N^2 D.
\end{equation}
By applying $C_N^2$ expressed in Eq. \eqref{eq: cnsss} in the SM, the structure function $D_\Phi $ is applicable for both a long-wave-dominated
spectrum of turbulence 
on scales below the correlation scale ($L$) and a short-wave-dominated spectrum on scales above the correlation scale ($l_0$). 
\footnote{Besides introducing the power-law spectrum in Fourier space, the structure function can be also derived by 
employing the real-space statistics. 
For instance, the rotation measure structure function calculated by using the correlation function within the inertial range of turbulence in 
\citet{LP15}
has the scaling consistent with that shown in 
Eq. \eqref{eq: sflp}  
(see equations (148) and (149) in \citealt{LP15}) in the case of a thick Faraday screen.}

The transverse separation across which the rms phase perturbation is equal to $1$ rad is defined as the diffractive length scale
(e.g., \citealt{Ric90}). By using Eq. \eqref{eq: sfspec}, it is expressed as
\begin{subnumcases}
   {r_\text{diff} \sim \label{eq: rdiff}}
   \Big(\pi r_e^2 \lambda^2 \text{SM} l_0^{\beta-4}\Big)^{-\frac{1}{2}},  ~~r_\text{diff}<l_0, \label{eq: rmina} \\
   \Big(\pi r_e^2 \lambda^2 \text{SM}\Big)^\frac{1}{2-\beta}, ~~~~~~~~r_\text{diff}>l_0. \label{eq: rminb}
\end{subnumcases}
In a particular case when $r_\text{diff}$ coincides with $l_0$, equaling $r_\text{diff}$ from the above equation and $l_0$ yields 
\begin{equation}
   \pi r_e^2 \lambda^2 \text{SM} l_0^{\beta-2} =1. 
\end{equation}
It means that the physical parameters involved in the scattering process should satisfy the condition 
\begin{equation}\label{eq: turpar1}
 \pi r_e^2 \lambda^2 \text{SM} l_0^{\beta-2} >1
\end{equation} 
for $r_\text{diff}$ to be smaller than $l_0$, and 
\begin{equation}\label{eq: turpar2}
  \pi r_e^2 \lambda^2 \text{SM} l_0^{\beta-2} <1
\end{equation}  
for $r_\text{diff}>l_0$ to be realized.
In terms of $r_\text{diff}$, $D_\Phi$ given by Eq. \eqref{eq: sfspec} can be written in the form
\begin{subnumcases}
 {D_\Phi =}
 \Big(\frac{r}{r_\text{diff}}\Big)^2, ~~~~~~~~~~~ r < l_0, \label{eq: dsfgau} \\
 \Big(\frac{r}{r_\text{diff}}\Big)^{\beta-2}, ~~~~~~~r>l_0.
\end{subnumcases}
In the presence of the inner scale of the density power spectrum, $D_\Phi$ exhibits a break in the slope at $r=l_0$ and steepens at smaller scales.
The quadratic scaling of $D_\Phi$ with $r$ at $r<l_0$
comes from the Gaussian distribution of density fluctuations $\exp(-k^2 l_0^2)$ below the inner scale (Eq. \eqref{eq: orisf}).

For the multi-path propagation in the strong scattering regime, $r_\text{diff}$ characterizes the coherent scale of the random 
phase fluctuations and the density perturbation on $r_\text{diff}$ dominates the scattering strength, with the 
angular and temporal broadening given by 
\citep{Ric90,Nara92}
\begin{equation}
   \theta_\text{sc} = \frac{\lambda}{2\pi r_\text{diff}}, 
\end{equation}
and
\begin{equation}\label{eq: tscpw}
  \tau_\text{sc} = \frac{D}{c} \theta_\text{sc}^2 = \frac{D\lambda^2}{4\pi^2 c} r_\text{diff}^{-2}.
\end{equation}
The above formulae pertain to the Galactic scattering medium, but should be modified when  
the scattering plasma is located at a cosmological distance. 
In the observer's frame, 
the wavelength is $\lambda_0 = \lambda (1+z_\text{q})$, where $z_\text{q}$ is the redshift of the scattering material.  
%\citep{Koa15}
%\begin{equation}
%   \theta_\text{sc,obs} =  \frac{D_\text{qp}}{D_\text{p}} \frac{\lambda_0}{2\pi r_\text{diff}},
%\end{equation}
By also taking into account the LOS weighting which depends on the location of the scattering material along the LOS
\citep{Gw93, Mac13},
the temporal broadening becomes
\begin{equation}\label{eq: obsw}
   W = \tau_\text{sc,obs} = (1+z_\text{q}) \frac{D_\text{eff}}{c} \theta_\text{sc}^2 = 
   \frac{D_\text{eff} \lambda_0^2}{ 4 \pi^2 c (1+z_\text{q})}  r_\text{diff}^{-2}.
\end{equation}
Here $D$ in Eq. \eqref{eq: tscpw} is replaced by the effective scattering distance $D_\text{eff} = D_\text{q}D_\text{qp}/D_\text{p}$,
with $D_\text{p}$, $D_\text{qp}$, and $D_q$ as the angular diameter distances from the observer to the source, from the source to the scattering medium,
and from the observer to the scattering medium. 
Accordingly, SM is also replaced with the weighted SM as adopted in 
\citet{Cor02},
\begin{equation}\label{eq: adjsm}
  \text{SM}  \sim  C_N^2 D_\text{eff}.
\end{equation}
$D_\text{eff}$ is comparable to $D_\text{q}$ in the case of Galactic scattering, and comparable to $D_\text{qp}$ when the scattering medium 
is close to the source. 
In both cases, $D_\text{eff}$ serves as a good approximation of the path length through the scattering region, and thus Eq. \eqref{eq: adjsm} is appropriate 
for estimating the actual SM. 
But we caution that for a thin scattering screen located somewhere between the source and the observer, its thickness, i.e., the path length 
that should be used for calculating SM, is in fact far smaller than the value of $D_\text{eff}$.

In combination with Eq. \eqref{eq: cnstep}, \eqref{eq: rdiff}, and \eqref{eq: adjsm}, 
the approximate expression of $W$ in the case of $\beta >3$ can be obtained from Eq. \eqref{eq: obsw}, 
\begin{subnumcases}
    {W \sim \label{eq: tscbst}}
  \frac{D_\text{eff}^2 r_e^2 \lambda_0^4}{4\pi c (1+z_\text{q})^3}  (\delta n_e)^2  \Big(\frac{l_0}{L}\Big)^{\beta-4} L^{-1},  \label{eq: scste}\\ \nonumber
  ~~~~~~~~~~~~~~~~~~~~~~~~~~~~~~~~~~~~~~~~~~~~~~~~~~~~~~~r_\text{diff} < l_0,  \\
  \frac{D_\text{eff}^\frac{\beta}{\beta-2} r_e^\frac{4}{\beta-2}  \lambda_0^\frac{2\beta}{\beta-2}  }{4\pi^\frac{2(\beta-3)}{\beta-2}   c  (1+z_\text{q})^\frac{\beta+2}{\beta-2}}   
  (\delta n_e)^\frac{4}{\beta-2} L^\frac{2(3-\beta)}{\beta-2}, \label{eq: scsteb}\\ \nonumber
  ~~~~~~~~~~~~~~~~~~~~~~~~~~~~~~~~~~~~~~~~~~~~~~~~~~~~~~~ r_\text{diff} > l_0.
\end{subnumcases}
The observationally measured wavelength dependence of the pulse width can make a distinction between the scenarios with $r_\text{diff}$ below 
or exceeding $l_0$, which, however, is limited by the insufficient accuracy of the current data 
\citep{LG14, Tho13}. 
Nevertheless, it is evident that in both situations $W$ decreases with increasing $L$. 
A given pulse width imposes a constraint on the outer scale of turbulence.
In particular, when $r_\text{diff} < l_0$, $W$ also decreases with increasing $l_0$. 
%The dependence of $W$ on $L$ and $l_0$ is expected as the turbulent power at a scale closer to $r_\text{diff}$ can more effectively enhance the diffractive 
%scattering. 
Moreover, in terms of the dispersion measure $\text{DM} = n_e D_\text{eff}$ of the scattering medium, where $n_e$ is the electron density averaged along 
the LOS passing through the scattering region, $W$ in Eq. \eqref{eq: tscbst} is rewritten as 
\begin{subnumcases}
    {W \sim }
  \frac{r_e^2 \lambda_0^4}{4\pi c (1+z_\text{q})^3} \Big(\frac{\delta n_e}{n_e}\Big)^2  \Big(\frac{l_0}{L}\Big)^{\beta-4} L^{-1} \text{DM}^2 ,  \label{eq: wdssta} \\ \nonumber
  ~~~~~~~~~~~~~~~~~~~~~~~~~~~~~~~~~~~~~~~~~~~~~~~~~~~~~~~r_\text{diff} < l_0,  \\
  \frac{r_e^\frac{4}{\beta-2}  \lambda_0^\frac{2\beta}{\beta-2}  }{4\pi^\frac{2(\beta-3)}{\beta-2}   c  (1+z_\text{q})^\frac{\beta+2}{\beta-2}}  
  \Big(\frac{\delta n_e}{n_e}\Big)^\frac{\beta}{\beta-2} \label{eq: wdsstb} \\ \nonumber
  (\delta n_e)^\frac{4-\beta}{\beta-2} L^\frac{2(3-\beta)}{\beta-2} \text{DM}^\frac{\beta}{\beta-2}, ~~~~~~~~~~~~ r_\text{diff} > l_0.
\end{subnumcases}

In the case of $\beta <3$, from Eq. \eqref{eq: cnshal}, \eqref{eq: rdiff}, \eqref{eq: obsw}, and \eqref{eq: adjsm}, $W$ can be estimated as 
\begin{subnumcases}
    {W \sim \label{eq: sctsharec}}
  \frac{D_\text{eff}^2 r_e^2 \lambda_0^4}{4\pi c (1+z_\text{q})^3}  (\delta n_e)^2   l_0^{-1}, ~~~~~~~~~~~ r_\text{diff} < l_0, \label{eq: scsha} \\
  \frac{D_\text{eff}^\frac{\beta}{\beta-2} r_e^\frac{4}{\beta-2}  \lambda_0^\frac{2\beta}{\beta-2} }{4 \pi^\frac{2(\beta-3)}{\beta-2} c (1+z_\text{q})^\frac{\beta+2}{\beta-2}}   
  (\delta n_e)^\frac{4}{\beta-2} l_0^\frac{2(3-\beta)}{\beta-2},  \label{eq: scshab}\\  \nonumber
  ~~~~~~~~~~~~~~~~~~~~~~~~~~~~~~~~~~~~~~~~~~~~~~~~~~~~ r_\text{diff} > l_0. 
\end{subnumcases}
Instead of $L$, $W$ in this case only places constraint on $l_0$. 
When $r_\text{diff} < l_0$, an excess of temporal broadening requires $l_0$ to be comparable to $r_\text{diff}$, 
so $l_0$ should be sufficiently small, while when $r_\text{diff} > l_0$, a larger $l_0$ is more favorable.
The relation between $W$ and DM can be also established 
\begin{subnumcases}
    {W \sim }
  \frac{ r_e^2 \lambda_0^4}{4\pi c (1+z_\text{q})^3}  \Big(\frac{\delta n_e}{n_e}\Big)^2   l_0^{-1} \text{DM}^2, ~ r_\text{diff} < l_0, \label{eq: wdssha} \\
  \frac{ r_e^\frac{4}{\beta-2}  \lambda_0^\frac{2\beta}{\beta-2} }{4 \pi^\frac{2(\beta-3)}{\beta-2} c (1+z_\text{q})^\frac{\beta+2}{\beta-2}}   
  \Big(\frac{\delta n_e}{n_e}\Big)^\frac{\beta}{\beta-2} \label{eq: wdsshb} \\ \nonumber
  (\delta n_e)^\frac{4-\beta}{\beta-2} l_0^\frac{2(3-\beta)}{\beta-2} \text{DM}^\frac{\beta}{\beta-2},  
  ~~~~~~~~~~~~ r_\text{diff} > l_0. 
\end{subnumcases}
By comparing Eq. \eqref{eq: wdssta}, \eqref{eq: wdsstb}, \eqref{eq: wdssha}, and \eqref{eq: wdsshb}, one can see that 
the dependence of $W$ on DM is determined by both the relation between $r_\text{diff}$ and $l_0$, and the spectral properties of density fluctuations. 
In general, 
$W$ increases more drastically with DM at a smaller $\beta$ in the case of $r_\text{diff}>l_0$, and has its mildest dependence on DM 
as $W \propto \text{DM}^2$ in the case of $r_\text{diff}<l_0$, irrespective of the value of $\beta$.
Also, the density perturbation $\delta n_e$ at $L$ for a long-wave-dominated spectrum of density fluctuations is close to $n_e$ averaged over a large scale, 
while $\delta n_e$ at $l_0$ for a short-wave-dominated spectrum can considerably exceed the background $n_e$ due to turbulent compression in shock-dominated flows. 
More exactly, following the power-law behavior, the ratio of the density perturbation at $l_0$ to that at $L$ when $\beta<3$ is 
\begin{equation}\label{eq: ratlgd}
    \frac{\delta n_e (l_0)}{\delta n_e (L)} = \frac{\delta n_e}{\delta n_e (L)} = \Big(\frac{L}{l_0}\Big)^\frac{3-\beta}{2}. 
\end{equation}
Therefore, with a higher density perturbation and a smaller scale $l_0$ instead of $L$ involved, 
a short-wave-dominated spectrum of density fluctuations provides much stronger scattering than a long-wave-dominated one when the DMs are the same.

\subsection{Applications in the IGM and the host galaxy ISM}\label{ssec: appl}

To elucidate the millisecond scattering tail observed for some FRBs 
\citep{Lor07, Tho13}, 
we next consider the IGM and the FRB host galaxy as two possible sources responsible for the scattering timescale.

(1) {\it Scattering in the IGM}

Growing observational evidence supports the presence of the IGM turbulence
(e.g.,  \citealt{Rau01, Zh04, Mei09, Lu10})
and the Kolmogorov type of turbulence in clusters of galaxies
\citep{Scu04, Ma04, Vo05}.
%The spectrum of intergalactic density fluctuations
%which can significantly deviate from that of turbulent velocity 
%is still poorly constrained. 
Supercomputer simulations show that the turbulent motions inside clusters of galaxies are subsonic, 
and are transonic or mildly supersonic in filaments
\citep{Ryu08},
which agrees with the observational detection of subsonic turbulence in e.g., the Coma cluster
\citep{Scu04}, 
the core of the Perseus cluster
\citep{Chu04}.
Down to small scales, theoretical studies suggest the existence of Alfv\'{e}nic turbulence with a spectrum dictated by 
the Kolmogorov scaling 
\citep{Sch06}, 
which is supported by the observed spectrum of magnetic energy in the core region of the Hydra cluster
\citep{Vo05}.
Based on these signatures obtained so far, the IGM turbulence is unlikely to be highly supersonic and thus unlikely
to possess a short-wave-dominated density spectrum, 
especially on scales small enough to be important for diffractive scattering.
Therefore, to
numerically evaluate the temporal broadening for propagation of radio waves through the diffuse IGM,
we consider a long-wave-dominated spectrum ($\beta>3$) of turbulent density 
and adopt the generally accepted Kolmogorov turbulence model with $\beta=11/3$.
Meanwhile, the choice of parameters should also be made to fulfill the conditions indicated by Eq. 
\eqref{eq: turpar1} and \eqref{eq: turpar2} in cases of $r_\text{diff} < l_0$ and $r_\text{diff} > l_0$, respectively. 
Inserting Eq. \eqref{eq: cnstep}, \eqref{eq: adjsm}, and $\beta=11/3$ into Eq. \eqref{eq: turpar1} and \eqref{eq: turpar2} yields
\begin{subequations}
\begin{align}
   L \Big(\frac{l_0}{L}\Big)^\frac{5}{3} > (\pi r_e^2 D_\text{eff} \lambda^2  (\delta n_e)^2)^{-1}, ~~~~ r_\text{diff} < l_0, \label{eq: lratco} \\
   L \Big(\frac{l_0}{L}\Big)^\frac{5}{3} < (\pi r_e^2 D_\text{eff} \lambda^2  (\delta n_e)^2)^{-1}, ~~~~ r_\text{diff} > l_0.
\end{align}
\end{subequations}
We now rewrite $W$ from Eq. \eqref{eq: tscbst} in terms of typical parameters for the IGM, 
\begin{subnumcases}
    {W \sim }
      \frac{0.065}{(1+z_\text{q})^3} \Big(\frac{D_\text{eff}}{1 \text{Gpc}}\Big)^2 \Big(\frac{\lambda_0}{1 \text{m}}\Big)^4  \\ \nonumber
      \Big(\frac{\delta n_e}{10^{-7}\text{cm}^{-3}}\Big)^2  
     \Big(\frac{l_0}{L}\Big)^{-\frac{1}{3}} \Big(\frac{L}{10^{-2} \text{pc}}\Big)^{-1} \text{ms} , ~~r_\text{diff} < l_0, \\
     \frac{4.9}{(1+z_\text{q})^{3.4}} \Big(\frac{D_\text{eff}}{1 \text{Gpc}}\Big)^{2.2} \Big(\frac{\lambda_0}{1 \text{m}}\Big)^{4.4}  \\ \nonumber
     \Big(\frac{\delta n_e}{10^{-7}\text{cm}^{-3}}\Big)^{2.4} 
     \Big(\frac{L}{10^{-2} \text{pc}}\Big)^{-0.8} \text{ms},  ~~~~~~~~~~ r_\text{diff} > l_0.
\end{subnumcases}
The value of $W$ at $r_\text{diff} < l_0$ depends on the disparity between $L$ and $l_0$, according to Eq. \eqref{eq: lratco}, 
which satisfies 
\begin{equation}
\begin{aligned}
    \frac{l_0}{L} & > 2.4\times10^{-6} (1+z_\text{q})^\frac{6}{5}
    \Big(\frac{D_\text{eff}}{1 \text{Gpc}}\Big)^{-\frac{3}{5}} \Big(\frac{\lambda_0}{1 \text{m}}\Big)^{-\frac{6}{5}} \\
  & ~~~~~ \Big(\frac{\delta n_e}{10^{-7}\text{cm}^{-3}}\Big)^{-\frac{6}{5}}  \Big(\frac{L}{10^{-2} \text{pc}}\Big)^{-\frac{3}{5}} .
\end{aligned}
\end{equation}
With the lower limit of $l_0/L$ in the above expression adopted, 
we get the same result in both cases that for a low-redshift source
the outer scale $L$ on the order of $10^{-2}$ pc can lead to 
the pulse duration of $\sim 5$ ms at $0.3$ GHz frequency ($\lambda=1$ m). 
The derived outer scale of turbulence seems unreasonably small compared with the expected 
injection scale ($> 100$ kpc) of turbulence induced by cluster mergers
\citep{Sub06}
or cosmological shocks 
\citep{Ryu08,Ry10}.
Also, as pointed out by 
\citet{LG14}, 
a serious difficulty is that such a small outer scale is accompanied by a turbulent heating rate at 
\begin{equation}
    \tau_\text{heat}^{-1} \sim \frac{c_s}{L} = 0.005 \Big(\frac{L}{10^{-2} \text{pc}}\Big)^{-1} \Big(\frac{T}{10^5 \text{K}}\Big)^\frac{1}{2} \text{yr}^{-1}, 
\end{equation}
where $c_s$ is the sound speed. 
The typical IGM temperature $T$ ranges from $10^5-10^7$ K
\citep{Ryu08, Byk08}.
The heating rate is so high that it is incompatible with the cooling rate which is comparable to the inverse Hubble time. 
With regards to a Kolmogorov cascade with the turbulent energy injected at a scale considerably larger than $\sim 10^{-2}$ pc, the resulting 
electron density fluctuations in the IGM make a negligible contribution to the observed temporal scattering.

Due to the high heating rate at small scales in the IGM, 
any small-scale density enhancement would be rapidly erased by the thermal streaming motions in the IGM 
\citep{Cor16}.
For this reason, the scenario in which the scattering medium is concentrated and localized in a thin layer in the IGM may not reflect the 
reality. Based on this questionable assumption, 
one tends to overestimate the contribution to the pulse broadening from the IGM.

(2) {\it Scattering in the host galaxy ISM}

In the multiphase ISM of the Galaxy, 
the distribution of the electron density fluctuations throughout the diffuse WIM is described by a Kolmogorov spectrum
\citep{Armstrong95, CL10}, 
but exhibits a much shallower spectrum in the supersonic turbulence prevalent in inner regions of the Galaxy
(e.g. \citealt{Laz09rev,HF12}).
By assuming that the host galaxy of an FRB is similar to the Galaxy and the general properties of turbulence are applicable,
we next attribute the strong scattering to propagation of radio waves through the ISM of the host galaxy
%Due to the lack of knowledge about the host galaxy medium, we assume the Milky Way as representative of a host galaxy for parameter normalization, 
and analyze the scattering effects from the Kolmogorov and short-wave-dominated density spectra, respectively.

We again start with the Kolmogorov power law of turbulence. 
The resulting $W$ from Eq. \eqref{eq: tscbst} is 
\begin{subnumcases}
    {W \sim \label{eq: ismkol}}
    \frac{0.0065}{(1+z_\text{q})^3} \Big(\frac{D_\text{eff}}{1 \text{kpc}}\Big)^2 \Big(\frac{\lambda_0}{1 \text{m}}\Big)^4  \label{eq: ismtsca}  \\ \nonumber
    \Big(\frac{\delta n_e}{10^{-2}\text{cm}^{-3}}\Big)^2  
     \Big(\frac{l_0}{L}\Big)^{-\frac{1}{3}} \Big(\frac{L}{10^{-3} \text{pc}}\Big)^{-1} \text{ms} , ~~~~~r_\text{diff} < l_0,    \\
    \frac{1.9}{(1+z_\text{q})^{3.4}} \Big(\frac{D_\text{eff}}{1 \text{kpc}}\Big)^{2.2} \Big(\frac{\lambda_0}{1 \text{m}}\Big)^{4.4}  \\ \nonumber
    \Big(\frac{\delta n_e}{10^{-2}\text{cm}^{-3}}\Big)^{2.4} 
     \Big(\frac{L}{10^{-3} \text{pc}}\Big)^{-0.8} \text{ms},  ~~~~~~~~~~~~~ r_\text{diff} > l_0. 
\end{subnumcases}
Here the normalization of $D_\text{eff}$ is assigned a typical galaxy size and $\delta n_e$ the electron density in diffuse ISM, i.e., $\delta n_e \sim n_e$. 
At $r_\text{diff}<l_0$, using Eq. \eqref{eq: lratco}, we have 
\begin{equation}
\begin{aligned}
    \frac{l_0}{L}& > 3.8\times10^{-8} (1+z_\text{q})^\frac{6}{5}
    \Big(\frac{D_\text{eff}}{1 \text{kpc}}\Big)^{-\frac{3}{5}} \Big(\frac{\lambda_0}{1 \text{m}}\Big)^{-\frac{6}{5}} \\
   &~~~~~ \Big(\frac{\delta n_e}{10^{-2}\text{cm}^{-3}}\Big)^{-\frac{6}{5}}  \Big(\frac{L}{10^{-3} \text{pc}}\Big)^{-\frac{3}{5}} .
\end{aligned}
\end{equation}
Substituting the lower limit of the ratio $l_0/L$ into Eq. \eqref{eq: ismtsca} yields the consistent result on the value of $W$ as in the case of $r_\text{diff}>l_0$. 
We see that $L$ inferred from the millisecond pulse broadening 
is far smaller than the injection scale of the turbulence throughout the Galactic WIM,  
which is suggested to be on the order of $\sim 100$ pc by measuring the spectra of interstellar density fluctuations
\citep{CL10, Armstrong95}.
It is also below the smaller outer scale of a few parsecs of the turbulence found in the Galactic spiral arms
\citep{MS96, Hav04}.
This heightens the challenge to interpreting the driving mechanism of the Kolmogorov turbulence as well as the cooling efficiency in the host galaxy.

A plausible solution is that a short-wave-dominated spectrum of electron density fluctuations 
which is extracted from the observations of the inner Galaxy also applies in the ISM of the host galaxy. 
As the density power spectrum becomes flat in supersonic turbulence, 
%The scatter broadening measurements of Galactic pulsars with high DMs 
%reveal significant departures from the Kolmogorov scaling 
%\citep{Loh01, Loh04, Bha04, Lew13, Lew15, Kri15, Cor16},
%which can be interpreted by a shallow spectrum of turbulent density field
%(Xu \& Zhang 2016, in preparation).
%The observed structure functions of rotation measure fluctuations at low Galactic latitudes also indicate a shallow density spectrum 
%\citep{Hav04, Hav08, XuZ16}.
if the turbulent ISM of the host galaxy that the LOS traverses through contains highly supersonic turbulent motions and as a result 
is characterized by numerous small-scale clumpy density structures, 
we expect that the spectrum of 
electron density fluctuations deviates from the Kolmogorov power law and has $\beta<3$.

In the above calculations, we assume that the volume filling factor $f$ of the scattering material is comparable to unity, 
which is valid for a long-wave-dominated density spectrum characterized by large-scale density fluctuations. 
For small-scale clumpy density structures described by a short-wave-dominated density spectrum, however, it is necessary to consider that 
only a fraction of volume is filled by the overdense regions and replace $\delta n_e$ with $\sqrt{f} \delta n_e$.
In the case of the Galactic ISM, the WIM phase where the Kolmogorov density spectrum is present has $f\sim25\%$. 
In contrast, the filling factors of the cold neutral medium and molecular clouds are as low as $1\%$ and $0.05\%$
\citep{Tie,HS13}.
In these colder and denser phases which only fill a small fraction of the volume, the short-wave-dominated density spectrum gives rise to 
small-scale density structures with the spatial profile of the density field characterized by peaks of mass as a result of strong shocks
(see figure 2 in \citealt{Kim05}).
Therefore the small-scale density structures created within these phases have an even smaller value of $f$.
Accordingly, we include the effect of a small filling factor in the case of a short-wave-dominated density spectrum, 
so as to reach a more realistic evaluation of the scattering produced by the supersonic turbulence in the host ISM.

In the case of $r_\text{diff}<l_0$, by inserting Eq. \eqref{eq: cnshal}, \eqref{eq: adjsm} into Eq. \eqref{eq: turpar1}, we find
\begin{equation}
\begin{aligned}
    l_0  & > \Big(\pi r_e^2 D_\text{eff} \lambda^2 f (\delta n_e)^2 \Big)^{-1}  \\
           & =  1.3\times10^7 (1+z_\text{q})^2 \Big(\frac{D_\text{eff}}{1 \text{kpc}}\Big)^{-1}
     \Big(\frac{\lambda_0}{1 \text{m}}\Big)^{-2} \\
           &~~~~~ \Big(\frac{f}{10^{-6}}\Big)^{-1} \Big(\frac{\delta n_e}{10^{-1}\text{cm}^{-3}}\Big)^{-2} \text{cm}.
\end{aligned}
\end{equation}
Given the parameters adopted in the above expression, the minimum $l_0$ is comparable to the 
inner scale of the density spectrum in the Galactic ISM inferred from observations 
%is as small as $\sim 10^6-10^8$ cm
\citep{Spa90, Armstrong95, Bha04}. 
By using a larger value of $l_0$ and 
substituting the normalization parameters into Eq. \eqref{eq: scsha} for a short-wave-dominated
spectrum of density fluctuations,  we get 
\begin{equation}\label{eq: schostsh}
\begin{aligned}
 & W  \sim \frac{6.5}{(1+z_\text{q})^3} \Big(\frac{D_\text{eff}}{1 \text{kpc}}\Big)^2 \Big(\frac{\lambda_0}{1 \text{m}}\Big)^4 \Big(\frac{f}{10^{-6}}\Big)
 \Big(\frac{\delta n_e}{10^{-1}\text{cm}^{-3}}\Big)^2  \\
 &~~~~~~~~~~     \Big(\frac{l_0}{10^{-10} \text{pc}}\Big)^{-1} \text{ms} .
\end{aligned}
\end{equation}
As $r_\text{diff}$ is below the inner scale of density power spectrum, 
the scaling presented in the above equation is independent of the spectral slope $\beta$ of density fluctuations.
It shows that clumps of electron density $0.1$ cm$^{-3}$
and size $10^{-10}$ pc ($\sim 10^8$ cm) which occupy a small fraction of the volume of the host galaxy 
would be adequate to produce the observed scattering delay.

Individual clumps of excess electrons have been included for modeling the Galactic distribution of electrons and 
scattering properties of Galactic ISM 
\citep{Pyn99, Cor03,Cor16}.
The clumpy component of the ionized plasma introduced in these studies are associated with discrete
H~{\sc ii} regions or supernova remnants with a characteristic scale of $\sim 1$ pc
\citep{Hav04}.
However, based on Eq. \eqref{eq: schostsh}
we note that the density fluctuations appearing on parsec scales, unless the local density is extraordinarily high, 
are unable to cause the intense scattering related with some FRBs. 
Differently, we consider much smaller-scale density structures corresponding to a short-wave-dominated
density spectrum with a sufficiently small inner scale. 
If the host galaxy medium is dominated by supersonic turbulence,
in accordance with the concentrated density distribution induced by shock compression,
the spectral form is dominated by the formation of small-scale density fluctuations 
and exhibits a rather shallow slope.

Compared with the above situation with $r_\text{diff}<l_0$ (Eq. \eqref{eq: schostsh}), 
the density spectrum in the case of $r_\text{diff}>l_0$ can lead to a significantly larger degree of scattering due to 
the stronger dependence of $W$ on the physical parameters involved (see Eq. \eqref{eq: scshab} and \eqref{eq: wdsshb}).
When the $\beta$ value can be determined,  
the scaling relations presented in Eq. \eqref{eq: scshab} (or Eq. \eqref{eq: wdsshb}) can be used to constrain the turbulence properties.
%THE FOLLOWING STATEMENT MAY BE EXCESSIVE
%Without any priori knowledge of the spectral slope, here we skip  
%detailed discussion on this scenario to avoid ambiguities in our calculations, 
%but remark that the scatter-broadened pulse observed for FRBs can be explained by 
%a shallow spectrum of density fluctuations in their host ISM. }
This small-scale properties of turbulent density can account for more pronounced scattering observed for some FRBs,  
and can also provide a plausible scattering source for the Galactic pulsars with high DMs
(Xu \& Zhang 2016, in preparation).
It implies that with similar properties to that of the Galaxy, the host galaxy is adequate to provide the observed scattering strength for an FRB.

(3) {\it Locations of scattering and dispersion}

Above results inform us that a long-wave-dominated power law spectrum, e.g., the Kolmogorov spectrum, of electron density fluctuations
with a reasonably large outer scale of turbulence in both the diffuse IGM and the host galaxy medium
are incapable of producing the millisecond scattering tail.
A short-wave-dominated electron density spectrum with $\beta<3$ 
from the ISM of the host galaxy can easily render the host galaxy a strong scatterer.
The excess fluctuation power at small scales characterized by a short-wave-dominated density spectrum gives rise to enhanced diffractive 
scattering and thus strong temporal broadening of a transient pulse.
%the diffractive effect, such as pulse broadening, is insensitive to large-scale density variations but is
%dominated by the density structures at sufficiently small scales comparable to $r_\text{diff}$.
%Thus a turbulence model resulting in a localized excess in electron density at a small scale is more favorable
%as the interpretation of  

A short-wave-dominated spectrum of density fluctuations in Galactic ISM can also produce the desired amount of scattering for FRBs. 
However, most of the known FRBs were discovered at high Galactic latitudes in directions through the WIM component of the ISM, 
where the turbulence is transonic
\citep{Haf99,Kim05,Hi08}
and the density fluctuations follow a Kolmogorov spectrum
\citep{Armstrong95, CL10}
with little scattering effect. 
Comparisons with the Galactic pulsars detected at comparable latitudes 
confirm the negligible Galactic contribution to the temporal broadening of FRBs 
\citep{Cor16}.
In fact, the heavy scattering from the supersonic turbulence that pervades the inner Galaxy prevents the detection of FRBs
\citep{Cor16}.

It is commonly accepted that the diffuse IGM makes unimportant contribution to scattering. 
Instead, intervening galactic halos along the LOS are appealed to for explaining the observed scattering
\citep{Yao16}.
Indeed, if the intervening ISM happens to be in a state of supersonic turbulence, 
and located close to us with a small reduction factor which depends on redshift, 
the intervening galaxy would dominate the scattering. 
However, we regard this scenario implausible because 
for a source at a cosmological distance, the probability for the LOS to intersect with an intervening galaxy is very low, e.g., 
$\leq 5\%$ within $z_q \sim 1.5$
\citep{Mac13, Roe69},
and the probability for the intervening ISM to be supersonically turbulent is further lower. 
This is in contradiction with the fact that around half of the known FRBs have detectable scattering tails
\citep{Pet16}.

After identifying the host galaxy medium as the most promising candidate for dominating the observed scattering, we see from Eq. \eqref{eq: schostsh} that 
\begin{equation}\label{eq: wdmfrb}
\begin{aligned}
 & W  \sim \frac{6.5}{(1+z_\text{q})^3}  \Big(\frac{\lambda_0}{1 \text{m}}\Big)^4 \Big(\frac{f}{10^{-6}}\Big)
 \Big(\frac{\delta n_e}{n_e}\Big)^2  \\
 &~~~~~~~~~~     \Big(\frac{l_0}{10^{-10} \text{pc}}\Big)^{-1} \Big(\frac{\text{DM}}{100~ \text{pc cm}^{-3}}\Big)^2  \text{ms} .
\end{aligned}
\end{equation}
The dependence of $W$ on DM 
is affected by the turbulence properties in the surrounding ISM of the source. 
Under the condition of a short-wave-dominated spectrum of density fluctuations, strong scattering does not entail large DM in the host medium.
As the Galactic contribution to the total DM is minor compared with its extragalactic component
\citep{Cor16},
the IGM is most likely the dominant location for the observed DMs of FRBs.

The FRB data exhibit considerable scatter around any modeled 
\citep{Ca16}
or fitted 
\citep{Yao16}
scattering time-DM relation. 
After considering an order of magnitude scatter similar to the case of Galactic pulsars 
\citep{Bha04}, 
one still cannot reach a satisfactory fit of the intergalactic scattering model to the FRB data
\citep{Ca16}.
As suggested in 
\citet{Ca16},
the LOS-dependent inhomogeneity in the Galactic ISM 
\citep{Jo98, Cor02}
may not apply to the IGM, which further poses difficulty for the IGM scattering scenario.
Besides, by plotting the scattering time vs. DM for high-Galactic latitude FRBs, 
\citet{Ka16}
claimed that no correlation between the two variables can be seen. 
More plausibly, scattering and dispersion are separately dominated by the host galaxy and the IGM. 
As shown above, 
the scattering time is largely affected by the turbulence properties (e.g., $\beta>3$ or $\beta<3$)
and scattering regimes ($r_\text{diff}<l_0$ or $r_\text{diff}>l_0$).
Therefore, the variation of the scattering time for FRBs can be attributable to the diverse interstellar environments of their host galaxies. 
From the observational point of view, it is also necessary to point out that 
the estimated scattering time is subject to effects such as the signal-to-noise ratio and limited temporal resolution due to dispersion smearing, 
leading to non-negligible uncertainties in the observationally measured scattering time-DM relation.

\section{Alternative models of electron density fluctuations}

Besides the turbulent cascade, 
different magnetic field structures associated with other processes such as plasma instabilities, fluctuation dynamo, can also induce electron density fluctuations 
at small scales in a compressible fluid. 
We next explore alternative models other than the power-law spectrum of density irregularities and their effects on the temporal broadening.

We first express the scattering time in a more general form, 
\begin{equation}
  \tau_{sc}  = \frac{D \lambda^2}{4\pi^2 c} r_\text{diff}^{-2} 
                  = \frac{D^2 r_e^2 \lambda^4}{4\pi c}  \frac{(\delta n_e(r_\text{diff}))^2}{ r_\text{diff}}, 
\end{equation}
from which Eq. \eqref{eq: tscbst} and \eqref{eq: sctsharec} can be recovered (see Appendix \ref{app}).
In a simple case when the fluctuating density $\delta n_e$ has a characteristic scale $d$, the above expression leads to 
\begin{equation}
  \tau_{sc}  = \frac{D^2 r_e^2 \lambda^4}{4\pi c}   \frac{(\delta n_e(d))^2}{d},
\end{equation}
which in the observer's frame is 
\begin{equation}\label{eq: tscsimpo}
  W  =   \frac{D_\text{eff}^2 r_e^2 \lambda_0^4}{4\pi c (1+z_\text{q})^3}   \frac{(\delta n_e(d))^2}{d}.
\end{equation}
In the strong scattering regime, the rms phase perturbation is greater than $1$ rad, i.e., $\sqrt{D_\Phi}>1$
(e.g. \citealt{Ric90,CoL91,LG14}).
Accordingly, we have a lower limit of $d$ at a given density perturbation 
\begin{equation}\label{eq: mimdo}
   d > (1+z_\text{q})^2 [\pi r_e^2 \lambda_0^2 D_\text{eff} (\delta n_e(d))^2 ]^{-1},
\end{equation}
or a lower limit of $\delta n_e(d)$ when $d$ is determined. 
\begin{equation}\label{eq: mimdeo}
    (\delta n_e(d))^2 >  (1+z_\text{q})^2 [\pi r_e^2 \lambda_0^2 D_\text{eff} d ]^{-1}.
\end{equation}
The derivations of the above equations are presented in Appendix \ref{app}. 
In the following analysis, we will apply these relations and the observational constraint on the scattering timescale
to investigate the scattering effect of other possibilities of density fluctuations.

\subsection{Electron density fluctuations arising from the mirror instability in the IGM}\label{eq: relnas}

For intergalactic plasmas, the ion collision frequency $\nu_{ii}$ is much lower than the cyclotron frequency $\Omega_i$, and accordingly, 
the mean free path of ions 
\citep{Braginskii:1965}
\begin{equation}\label{eq: mfpio}
\begin{aligned}
     \lambda_\text{mfp} &= \frac{ v_\text{th,i}}{ \nu_{ii}} = \frac{3\sqrt{2}(k_B T)^{2}}{4 \sqrt{\pi} \ln\Lambda e^4 n_i} \\
   & =  2.15\times10^{21} \Big(\frac{\ln \Lambda}{10}\Big)^{-1}  \Big(\frac{T}{10^5 \text{K}}\Big)^2  \Big(\frac{n_i}{10^{-7}\text{cm}^{-3}}\Big)^{-1} \text{cm}
\end{aligned}
\end{equation}
is significantly larger than the ion gyroradius
\begin{equation}\label{eq: igyros}
\begin{aligned}
   l_i &= \frac{v_\text{th,i}}{\Omega_i}  = \frac{v_\text{th,i}m_i c}{eB} \\
   &=4.2\times10^9 \Big(\frac{T}{10^5 \text{K}}\Big)^\frac{1}{2} \Big(\frac{B}{0.1 \mu \text{G}}\Big)^{-1} \text{cm}, 
\end{aligned}
\end{equation}
where $v_\text{th,i} = \sqrt{2k_B T/ m_i}$ is the ion thermal speed, and 
$k_B$, $\ln \Lambda$, $n_i$, $c$ are the Boltzmann constant, Coulomb logarithm, ion number density, and speed of light.
The magnetic field strength $B$ is taken as the inferred value from the Faraday rotation measures of polarized extragalactic sources
\citep{Ryu98, Xu06}.
We also treat the IGM as a fully ionized hydrogen plasma, so ions have the same charge $e$ and mass $m_i = m_H$ as protons. 

The weakly collisional and magnetized IGM is subject to firehose and mirror instabilities driven by pressure anisotropies with respect to the local magnetic field direction
\citep{Fab94,Car02,Sch05,Rin15}.
The instability growth rate increases with wave numbers, resulting in 
fluctuating magnetic fields peaking at a plasma micro-scale comparable to the ion gyro-scale $l_i$
\citep{Sch06}. 
The compressive mirror instability induces variations in density which are anti-correlated with the magnetic field variations. 
The fluctuations in density and magnetic field are related as
\citep{Ha80},
\begin{equation}
     \frac{\delta n_e}{n_{e}} \sim \frac{\delta B}{B}, 
\end{equation}
where $\delta n_e$, $\delta B$ and $n_{e}$, $B$ are the fluctuating and uniform components of electron density and magnetic field strength, respectively.
If the density perturbation $\delta n_e(d)$ at $d = l_i \sim 4.2\times10^9$ cm (Eq. \eqref{eq: igyros})
is sufficient to account for strong scattering, Eq. \eqref{eq: mimdeo} sets 
the lower limit of density perturbation at $d$,
\begin{equation} \label{eq: esdmim}
\begin{aligned}
 &   \delta n_e(d)  >  5.5 \times10^{-9} (1+z_\text{q}) 
    \Big(\frac{D_\text{eff}}{1\text{Gpc}}\Big)^{-\frac{1}{2}} \Big(\frac{\lambda_0}{1\text{m}}\Big)^{-1} \\
 &~~~~~~~~~~~~~~~~  \Big(\frac{T}{10^5 \text{K}}\Big)^{-\frac{1}{4}} \Big(\frac{B}{0.1 \mu \text{G}}\Big)^\frac{1}{2} \text{cm}^{-3}.
\end{aligned}
\end{equation}
Inserting the above expression and Eq. \eqref{eq: igyros} into Eq. \eqref{eq: tscsimpo} results in 
\begin{equation}
    W > \frac{1.5\times10^3}{1+z_\text{q}} \Big(\frac{D_\text{eff}}{1\text{Gpc}}\Big)  \Big(\frac{\lambda_0}{1\text{m}}\Big)^2 
    \Big(\frac{T}{10^5 \text{K}}\Big)^{-1} \Big(\frac{B}{0.1 \mu \text{G}}\Big)^{2}   \text{ms}. 
\end{equation}
The predicted timescale is obviously inconsistent with the observed FRB pulses with millisecond or shorter durations.   
To accommodate the observations, 
the saturated amplitude of the density fluctuations and the associated magnetic fluctuations generated by plasma instabilities should remain 
at a marginal level, so that the strong scattering cannot be realized. 
We can see from Eq. \eqref{eq: esdmim} that by adopting  
an average electron density as $n_e = 10^{-7}$ cm$^{-3}$, a conservative estimate of the magnetic field and density perturbations near $l_i$ is 
\begin{equation}
      \frac{\delta B}{B}  \sim  \frac{\delta n_e}{n_{e}} < \frac{5.5 \times10^{-9} \text{cm}^{-3}}{10^{-7} \text{cm}^{-3}} =0.055. 
\end{equation}
This result suggests that although the micro-plasma instabilities have a fast growth rate in comparison with the large-scale turbulent motions, they 
are mostly suppressed over the fluid timescale. 
As demonstrated by earlier works, the enhanced particle scattering originating from the plasma instabilities can effectively relax the pressure anisotropy and 
increase the collision rate. 
As a result, both the turbulent cascade over small scales and efficient magnetic field amplification can be facilitated
\citep{LB06, Rei14}.
This naturally explains the magnetization and turbulent motions in the IGM inferred from the observations
(e.g., \citealt{Ryu08}).
By taking into account the relaxation effect of pressure anisotropy, the collisionless MHD simulations carried out by 
\citet{Rei14}
exhibit the statistical properties of turbulence similar to that of collisional MHD turbulence, which justifies a collisional-MHD description
of collisionless plasmas at the intracluster medium (and IGM) conditions. 
The observed pulse widths of transient radio sources at cosmological distances, like the FRBs, offer a strong argument supporting the above picture of the 
IGM turbulence, 
whereas the model of nonlinear evolution of the plasma instabilities with a secular growth of small-scale magnetic field fluctuations to large amplitudes, $\delta B / B \sim 1$, 
is disfavored
\citep{Sch08, Rin15}.

\subsection{Electron density fluctuations arising from a folded structure of magnetic fields in the IGM}\label{ssec:igmf}

Corresponding to the large mean free path of ions in the IGM, the viscosity parallel to magnetic field lines is 
\begin{equation}\label{eq: vispara}
\begin{aligned}
    \nu_i &= \lambda_\text{mfp} v_\text{th,i} =  \frac{3(k_B T)^\frac{5}{2}}{2 \sqrt{\pi} \ln\Lambda \sqrt{m_i} e^4 n_i} \\
 & = 8.7\times10^{27} \Big(\frac{\ln \Lambda}{10}\Big)^{-1}  \Big(\frac{T}{10^5 \text{K}}\Big)^\frac{5}{2}  \Big(\frac{n_i}{10^{-7}\text{cm}^{-3}}\Big)^{-1} \text{cm}^2 \text{s}^{-1}.
\end{aligned}
\end{equation}
It damps the turbulent cascade at a large viscous scale, which can be obtained by equaling the turbulent cascading rate 
\begin{equation}
   \tau_\text{cas}^{-1} =\frac{v_l}{l} =k^\frac{2}{3}L^{-\frac{1}{3}} V_L
\end{equation}
with the viscous damping rate $k^2 \nu_i$. Here we use the Kolmogorov scaling, where $v_l$ is the turbulent velocity at scale $l$, 
$V_L$ is the turbulent velocity at the injection scale $L$, and $k = 1/l$ is the wavenumber.  
The viscous scale calculated by using the parallel viscosity is 
\begin{equation}\label{eq: lincig}
\begin{aligned}
   l_0 &=  L^\frac{1}{4} V_L^{-\frac{3}{4}} \nu_i^\frac{3}{4} \\
    & = 3.78\times10^{21} \Big(\frac{L}{100 \text{kpc}}\Big)^\frac{1}{4} \Big(\frac{V_L}{100 \text{km s}^{-1} }\Big)^{-\frac{3}{4}} \\
  & ~~~~~~  \Big(\frac{\ln \Lambda}{10}\Big)^{-\frac{3}{4}}  \Big(\frac{T}{10^5 \text{K}}\Big)^\frac{15}{8}  \Big(\frac{n_i}{10^{-7}\text{cm}^{-3}}\Big)^{-\frac{3}{4}} \text{cm}.
\end{aligned}
\end{equation}
The viscous-scale eddies are responsible for the random stretching of magnetic field lines that drives an exponential growth of the initially weak magnetic energy at a rate 
equal to the viscous-eddy turnover rate. 
As mentioned in Section \ref{eq: relnas}, 
the particle scattering in the presence of the plasma instabilities makes the effective parallel viscosity sufficiently small, 
and thus the corresponding dynamo growth rate becomes fast
\citep{Sch06,Rei14}, 
so that the kinematic dynamo process can be efficient enough to generate strong magnetic fields within the cluster lifetime.

In addition, the ordinary Spitzer resistivity in the IGM is negligibly small
\citep{Spit56},
\begin{equation}\label{eq: spiresis}
\begin{aligned}
 \eta &= \frac{c^2 \sqrt{m_e} e^2 \ln \Lambda}{4 (k_B T)^\frac{3}{2}}  \\
 &= 3.05\times10^5 \Big(\frac{\ln\Lambda}{10}\Big) \Big(\frac{T}{10^5 \text{K}}\Big)^{-\frac{3}{2}} \text{cm}^2 \text{s}^{-1}.
\end{aligned}
\end{equation}
Thus the magnetic Prandtl number $P_m = \nu_i / \eta \sim 10^{22}$ (Eq. \eqref{eq: vispara} and \eqref{eq: spiresis}) in the IGM is high,
and magnetic fluctuations can be developed in the viscosity damped regime of MHD turbulence
\citep{CLV_newregime, CLV03, LVC04}.
During the dynamo growth of magnetic energy, the stretched magnetic fields form a folded structure in the sub-viscous range, 
with the field variation along the field lines at the viscous scale $l_0$ and the field direction reversal at the resistive scale
\citep{Sch04,GS06,Laz07, Bra15}.
The folded magnetic fields compress gas into dense sheet-like structures. 
Such dense sheets have been invoked to explain the formation of 
the small ionized and neutral structures (SINS) in the partially ionized ISM 
\citep{Die76,Hei97,Sta04}
by
\citet{Laz07}, 
and is also proposed as the source of 
extreme diffractive scattering in the Galactic center by 
\citet{GS06}.

However,
as regards the fully ionized IGM environment, the persistence of the folded structure of magnetic fields is speculative. 
First, the folded structure especially its curved part is unstable to the plasma instabilities and the resulting thickness of the fold can be much larger 
than the resistive scale
\citep{Sch05}.
Moreover, not only the parallel viscosity can be effectively reduced, 
the viscosity perpendicular to magnetic field lines also substantially decreases with increasing field strength
\citep{Sim55},
\begin{equation}
\begin{aligned}
    \nu_{i,\perp} &= \frac{3k_BT \nu_{ii}}{10 \Omega_i^2 m_i} 
    = 5.1\times10^3 \Big(\frac{\ln \Lambda}{10}\Big)  \Big(\frac{T}{10^5 \text{K}}\Big)^{-\frac{1}{2}}  \\
 &~~~~~~~~~~~~~~  \Big(\frac{n_i}{10^{-7}\text{cm}^{-3}}\Big) \Big(\frac{B}{0.1 \mu \text{G}}\Big)^{-2} \text{cm}^2 \text{s}^{-1}. 
\end{aligned}
\end{equation}
It implies that the turbulent motions perpendicular to magnetic field lines are undamped at the viscous scale $l_0$ (Eq. \eqref{eq: lincig})
derived from the parallel viscosity
and can initiate a cascade of Alfv\'{e}nic turbulence at smaller scales down to the cutoff scale determined by the much smaller 
perpendicular viscosity, 
which tends to violate the preservation of the folded structure of magnetic fields on scales below $l_0$.

In the following analysis, 
%in disregard of the reduction of both parallel viscosity caused by particle scattering and perpendicular viscosity due to magnetic field growth, 
we nevertheless presume that 
the magnetic fields appear in folds with undetermined thickness at scales below $l_0$, and the local magnetic perturbation is determined by the 
equilibrium between the turbulent energy at $l_0$ and the magnetic-fluctuation energy, 
\begin{equation}\label{eq: delbvis}
    \delta B = \sqrt{4\pi\rho_{i}} v_0 = \sqrt{4\pi\rho_{i}} V_L \Big(\frac{l_0}{L}\Big)^\frac{1}{3},
\end{equation}
where $\rho_{i} = m_H n_i $ is the average mass density of ions. 
In pressure equilibrium, the density perturbation across the sheet of folded fields
is approximately given by the ratio between the local magnetic and gas pressure
\citep{Laz07}, 
\begin{equation}\label{eq: nnp}
     \frac{\delta n_e}{n_{e}} \sim \frac{P_B}{P_g} , 
\end{equation}
with the magnetic pressure $P_B = (\delta B) ^2/8\pi$, and the thermal pressure 
\begin{equation}\label{eq: rapp}
  P_g = P_i+P_e = n_i k_B T + n_e k_B T = 2 n_i k_B T, 
\end{equation}
where the number density of ions $n_i$ and electrons $n_e$ are equal. 
Therefore we can get (Eq. \eqref{eq: lincig}, \eqref{eq: delbvis}, \eqref{eq: nnp}, \eqref{eq: rapp})
\begin{equation}\label{eq: depig}
\begin{aligned}
    \frac{\delta n_e}{n_e} &= \frac{(\delta B) ^2}{16\pi n_i k_B T } = \frac{m_i V_L^2 l_0^\frac{2}{3}}{ 4 k_B T L^\frac{2}{3}}  \\
    &= 0.16 \Big(\frac{L}{100 \text{kpc}}\Big)^{-\frac{1}{2}} \Big(\frac{V_L}{100 \text{km s}^{-1} }\Big)^\frac{3}{2} \\
  & ~~~~~~  \Big(\frac{\ln \Lambda}{10}\Big)^{-\frac{1}{2}}  \Big(\frac{T}{10^5 \text{K}}\Big)^\frac{1}{4}  \Big(\frac{n_i}{10^{-7}\text{cm}^{-3}}\Big)^{-\frac{1}{2}} .
\end{aligned}
\end{equation}
By taking $\delta n_e (d) / n_e \sim  0.16$ from above expression and $n_e = 10^{-7}$ cm$^{-3}$, the condition of strong scattering requires (Eq. \eqref{eq: mimdo})
\begin{equation}
    d > 5.0 \times 10^8 (1+z_\text{q})^2 \Big(\frac{D_\text{eff}}{1 \text{Gpc}}\Big)^{-1} \Big(\frac{\lambda_0}{1\text{m}}\Big)^{-2} \text{cm},
\end{equation}
with the lower limit smaller than $l_i$ (Eq. \eqref{eq: igyros}). 
It implies that the density perturbation we adopt for the folded structure at any sub-viscous scale can contribute to strong scattering. 
We have demonstrated in Section \ref{ssec: appl} that the intergalactic scattering is likely weak.
Therefore, in accordance with the observationally determined scattering timescale $W$, 
the lower limit of the characteristic sheet thickness is set by (Eq. \eqref{eq: tscsimpo}), 
\begin{equation}\label{eq: llsth}
\begin{aligned}
    d &> \frac{D_\text{eff}^2 r_e^2 \lambda_0^4}{4\pi c (1+z_\text{q})^3}   \frac{(\delta n_e(d))^2}{W} \\
    &= \frac{5.2\times10^{13}}{(1+z_\text{q})^3} \Big(\frac{D_\text{eff}}{1 \text{Gpc}}\Big)^{2} \Big(\frac{\lambda_0}{1\text{m}}\Big)^{4} \Big(\frac{W}{1 \text{ms}}\Big)^{-1} \text{cm}.
\end{aligned}
\end{equation}
As expected, it is larger than the resistive scale, which can be calculated from Eq. \eqref{eq: vispara}, \eqref{eq: lincig}, and \eqref{eq: spiresis},
\begin{equation}
\begin{aligned}
   l_R = l_0 P_m^{-\frac{1}{2}} &= 2.2\times10^{10} \Big(\frac{L}{100 \text{kpc}}\Big)^\frac{1}{4} \Big(\frac{V_L}{100 \text{km s}^{-1}}\Big)^{-\frac{3}{4}} \\
         & \Big(\frac{\ln \Lambda}{10}\Big)^\frac{1}{4}  \Big(\frac{T}{10^5 \text{K}}\Big)^{-\frac{1}{8}} 
                          \Big(\frac{n_i}{10^{-7} \text{cm}^{-3}}\Big)^{-\frac{1}{4}} \text{cm}.
\end{aligned}
\end{equation}

For the small-scale folded magnetic fields generated by fluctuation dynamo, besides the geometrical structure that is related with the scattering effects on 
radiation propagation, in terms of one-dimensional magnetic energy spectrum in the viscosity-dominated regime,
a distinctive spectral slope of $k^{-1}$ has been analytically derived by 
\citet{LVC04}
and numerically confirmed by 
\citet{CLV_newregime}.
The detection of such a spectral index and comparison between the measured spectral cutoff scale and the lower limit of sheet thickness in Eq. \eqref{eq: llsth}
can verify the existence of the folded magnetic fields and 
provide more definite information on the properties of the viscosity damped regime of turbulence.

\subsection{Electron density fluctuations arising from a folded structure of magnetic fields in the host galaxy medium }\label{ssec:ismf}

The folded structure of magnetic fields in the sub-viscous range of turbulence can also be present in the ISM of the host galaxy. 
We next follow the similar calculations as shown above, but use the 
environment parameters for the Galactic WIM 
\citep{MO77}, 
which accounts for most of the ionized gas within the Galactic ISM 
\citep{Haf09}
and is taken as an example of the fully ionized phase of the host galaxy medium.

Given the parallel viscosity (Eq. \eqref{eq: vispara})
\begin{equation}
  \nu_i = 1.6\times10^{19} \Big(\frac{\ln \Lambda}{10}\Big)^{-1} \Big(\frac{T}{8000 \text{K}}\Big)^\frac{5}{2} 
       \Big(\frac{n_i}{0.1\text{cm}^{-3}}\Big)^{-1} \text{cm}^{2} \text{s}^{-1},
\end{equation}
and the Spitzer resistivity (Eq. \eqref{eq: spiresis})
\begin{equation}
   \eta =1.3\times10^7 \Big(\frac{\ln \Lambda}{10}\Big) \Big(\frac{T}{8000 \text{K}}\Big)^{-\frac{3}{2}} \text{cm}^{2} \text{s}^{-1},
\end{equation}
the WIM phase has a large $P_m$, 
\begin{equation}\label{eq: pm}
\begin{aligned}
   P_m &= \frac{\nu_i}{\eta} \\
   &= 1.2 \times10^{12} \Big(\frac{\ln \Lambda}{10}\Big)^{-2} \Big(\frac{T}{8000 \text{K}}\Big)^4  \Big(\frac{n_i}{0.1\text{cm}^{-3}}\Big)^{-1}. 
\end{aligned}
\end{equation}
The resulting resistive scale 
\begin{equation}\label{eq: reschome}
\begin{aligned}
   l_R = l_0 P_m^{-\frac{1}{2}} &= 7.2\times10^8 \Big(\frac{L}{30 \text{pc}}\Big)^\frac{1}{4} \Big(\frac{V_L}{10 \text{km s}^{-1}}\Big)^{-\frac{3}{4}} \\
         & \Big(\frac{\ln \Lambda}{10}\Big)^\frac{1}{4}  \Big(\frac{T}{8000 \text{K}}\Big)^{-\frac{1}{8}} 
                          \Big(\frac{n_i}{0.1\text{cm}^{-3}}\Big)^{-\frac{1}{4}} \text{cm}
\end{aligned}
\end{equation}
is smaller than the ion mean free path (Eq. \eqref{eq: mfpio})
\begin{equation}
     \lambda_\text{mfp} =  1.4\times10^{13} \Big(\frac{\ln \Lambda}{10}\Big)^{-1}  \Big(\frac{T}{8000 \text{K}}\Big)^2  \Big(\frac{n_i}{0.1 \text{cm}^{-3}}\Big)^{-1} \text{cm},
\end{equation}
and thus falls in the collisionless regime. 
Similar to the IGM plasma, the folded structure of magnetic fields can be significantly affected by the plasma instabilities and turbulent cascade at small scales. 
Nevertheless, to seek the possibility of enhanced scattering introduced by different structures of magnetic fields arising in the host galaxy medium, 
we suppose that the folded fields survive at scales below the viscous scale $l_0$ (Eq. \eqref{eq: lincig}), 
\begin{equation}
\begin{aligned}
   l_0  & =7.8\times10^{14}  \Big(\frac{L}{30 \text{pc}}\Big)^{\frac{1}{4}} \Big(\frac{V_L}{10 \text{km s}^{-1}}\Big)^{-\frac{3}{4}} \\
         &~~~~~~ \Big(\frac{\ln \Lambda}{10}\Big)^{-\frac{3}{4}} \Big(\frac{T}{8000 \text{K}}\Big)^{\frac{15}{8}} 
                          \Big(\frac{n_i}{0.1\text{cm}^{-3}}\Big)^{-\frac{3}{4}} \text{cm}.
\end{aligned}
\end{equation}
In the case of the WIM, 
the turbulent cascade along the long-wave-dominated Kolmogorov spectrum over an extended inertial range leads to small turbulent fluctuations at $l_0$. 
So the corresponding density perturbation given by Eq. \eqref{eq: depig} is relatively small, 
\begin{equation}
\begin{aligned}
    \frac{\delta n_e}{n_e} &= \frac{m_i V_L^2 l_0^\frac{2}{3}}{ 4 k_B T L^\frac{2}{3}}  \\
    &= 1.6\times10^{-4} \Big(\frac{L}{30 \text{pc}}\Big)^{-\frac{1}{2}} \Big(\frac{V_L}{10 \text{km s}^{-1} }\Big)^\frac{3}{2} \\
  & ~~~~~~  \Big(\frac{\ln \Lambda}{10}\Big)^{-\frac{1}{2}}  \Big(\frac{T}{8000 \text{K}}\Big)^\frac{1}{4}  \Big(\frac{n_i}{0.1\text{cm}^{-3}}\Big)^{-\frac{1}{2}} .
\end{aligned}
\end{equation}
It follows that to fulfill the strong scattering condition, the characteristic scale of the density fluctuations should be sufficiently large (Eq. \eqref{eq: mimdo}), 
\begin{equation}
    d > 5.3\times10^8 (1+z_\text{q})^2 \Big(\frac{D_\text{eff}}{1 \text{kpc}}\Big)^{-1} \Big(\frac{\lambda_0}{1\text{m}}\Big)^{-2} \text{cm}, 
\end{equation}
where $\delta n_e (d) / n_e \sim 1.6\times10^{-4}$ and $n_e = 0.1$ cm$^{-3}$ are used. 
But in the meantime, as the density perturbation is rather weak, only with a small value of $d$ can the millisecond pulse duration be reached (Eq. \eqref{eq: tscsimpo})
\begin{equation}
\begin{aligned}
    d \leq \frac{4.9\times10^7}{(1+z_\text{q})^3}  \Big(\frac{D_\text{eff}}{1 \text{kpc}}\Big)^{2} \Big(\frac{\lambda_0}{1\text{m}}\Big)^{4} \Big(\frac{W}{1 \text{ms}}\Big)^{-1} \text{cm}. 
\end{aligned}
\end{equation}
The thickness of the sheet-like structure in density field is expected to be larger than $l_R$ (Eq. \eqref{eq: reschome})
due to the effect of plasma instabilities 
\citep{Sch05}, 
and thus larger than the value indicated from the above equation, 
leading to insignificant pulse broadening.

This result shows that the density fluctuations induced by the folded structure of magnetic fields in the WIM-like environment 
are inadequate to render the host galaxy a strong scatterer. 
It has been suggested earlier by 
\citet{GS06}
that the large density contrast associated with the folded fields suffices for interpreting the extreme scattering of radio waves taking place in the Galactic center. 
Besides different environment parameters employed, as the major difference between our analysis and their work, 
we use the local magnetic field fluctuations with the magnetic energy equal to the turbulent energy at the viscous scale in deriving the density perturbation, 
rather than the magnetic field coherent on the scale of the largest turbulent eddy taken in 
\citet{GS06}, 
which has a much stronger strength than the perturbed field on the scale of the smallest eddy. 
The scenario described in 
\citet{GS06}
can be realized when the forcing scale of turbulence is comparable to the viscous scale and the inertial range of turbulence is absent. 
Otherwise the folded fields only emerge in the sub-viscous region with larger-scale magnetic perturbations irrelevant in determining the local density structure.

It is also necessary to point out that we use the isotropic Kolmogorov scaling for analytical simplicity 
in Section \ref{ssec:igmf} and \ref{ssec:ismf}. But in fact, as the magnetic field becomes dynamically 
important, anisotropic MHD turbulence develops with the turbulent eddies more elongated along the local magnetic field direction toward smaller scales. 
Then the 
\citet{GS95} 
scaling applies as a more appropriate description of the relation between the parallel and perpendicular scales with respect to the local magnetic field. 
If one takes into account the effect of turbulence anisotropy in the above calculations, the viscous damping rate $k^2 \nu_i$ is replaced by 
$k_\|^2 \nu_i$, and the latter is relatively small. 
Here $k_\|$ and $k_\perp$ are the parallel and perpendicular components of wavevector $\bm{k}$. 
Accordingly, the viscous scale is shifted downward and the corresponding density fluctuations are further reduced (Eq. \eqref{eq: depig}), leading to a 
less important contribution of the plasma sheets in scatter broadening.

\section{Discussion and conclusions}

We analyzed various models of electron density fluctuations and examined their effects on broadening FRB pulse widths. 
Different from earlier studies
(e.g., \citealt{Mac13, LG14})
where the Kolmogorov turbulence is conventionally adopted for describing the spatial power spectrum of density fluctuations, 
our study is devoted to a general form of the density spectrum, as well as other density structures induced by 
physical processes including plasma instabilities and fluctuation dynamo in both the IGM and ISM of the host galaxy.

\citet{Mac13} 
evaluated the strength of scattering in the IGM by assuming a Kolmogorov spectrum and 
a sufficiently low outer scale of turbulence. 
Our calculation under similar turbulence conditions yields detectable intergalactic scattering. 
We disfavor this picture
because as pointed out by 
\citet{LG14}, 
an outer scale smaller than $\sim 10^{24}$ cm entails too large turbulent heating rate to be compatible with the cooling rate in the realistic IGM. 
\citet{Yao16} 
suggested the importance of the IGM in both dispersion and scattering of FRBs and empirically determined a flat DM-dependence 
$\propto\text{DM}^{1.3}$ of the scattering timescale, 
which to our knowledge is inconsistent with the predictions of existing scattering theories. 
Furthermore, when confronted to the observational data of known FRBs, non-monotonic dependence of pulse widths on DMs is obviously seen 
\citep{Ka16, Kat16}, e.g., FRB 110703 has larger DM but shorter scattering timescale in comparison with FRB 110220 \citep{Tho13}. 
The considerable scatter around any single $W$-DM relation can be hardly interpreted as 
sightline-to-sightline scatter since the probability of encountering an intervening galaxy along the LOS is quite low 
\citep{Mac13}. 
An alternative scenario that the host galaxy dominates both dispersion and scattering was raised in
\citet{Cor16}
(see also \citealt{Han15}).
Their analysis was restricted to the Kolmogorov turbulence model and based on a specific relation between the broadening time and DM, $W \propto \text{DM}^2$,
which corresponds to Eq. \eqref{eq: wdssta} at $\beta =11/3$ and $r_\text{diff}<l_0$ in this work.
Our general discussion on the spectral properties of density fluctuations overcomes this limitation and enables us to gain new physical insight. 
We find that a short-wave-dominated spectrum of turbulent density in the host galaxy medium provides a plausible explanation of the pulse broadening of FRBs.
A single relation between the scattering and dispersion in the host galaxies for all FRBs is inappropriate because of the widely diverse turbulence properties in 
different host galaxies. 
The strong scattering effect can naturally arise as a consequence of a short-wave-dominated density spectrum 
and in the meantime the host-galaxy component of the total DM is small,
supportive of the dominant intergalactic contribution to dispersion and cosmological distances of FRBs.

A short-wave-dominated spectrum of density fluctuations is commonly observed in the inner Galaxy where the turbulent flows are 
highly supersonic and shock-dominated
\citep{Laz09rev, HF12, Fal14}. 
The turbulent energy is predominantly injected by stellar sources such as stellar winds and protostellar outflows, indicative of active star formation
\citep{Hav08}.
If an FRB resides in the center region of a galaxy with intense ongoing star formation where the 
power spectrum of density field becomes flat,
evident temporal broadening independent of the inclination angle of the host galaxy is expected. 
We caution that the situation regarding FRBs with discernible scattering tails is complicated 
by the fact that the observed pulse width can contain both the host galaxy component and the intrinsic one.
Therefore, extra care is needed when the pulse width is used as a discriminator between different progenitor models
\citep{Kea16}.

Among the diverse FRB progenitor models, some are indicative of rich and turbulent ISM environment with intense star formation. 
The discovery of repeating bursts from FRB 121102 
\citep{spitler16} 
supports an origin of young neutron stars, from which giant radio pulses may be sporadically produced
\citep{cordes16,connor16,Ly16}. 
These young neutron stars are likely to be found in star-forming regions where the requirement to produce strong scatter broadening can be easily met. 
The magnetar giant flare model 
\citep{Pop13,Kul14,Ka16}
also relates FRBs with young neutron stars, which mark the star-forming regions of galaxies.
For other repeating FRB models 
(e.g., \citealt{dai16, Gu16}),
scattering effect can also manifest to the observer
if their preferential environment is characterized by a high star-formation rate.
As for the non-repeating FRBs with distinct cosmological origins, the blitzar model 
\citep{falcke14,zhang14} 
invokes delayed collapse of a supra-massive neutron star to a black hole after it loses centrifugal support, with a timescale ranging from minutes 
\citep{zhang14} 
to thousands of years 
\citep{falcke14} 
after the birth of the neutron star. 
Plausibly, if the supra-massive neutron star comes from collapse of a massive star, this model is also related with star formation activity and satisfies the external condition for pulse broadening. 
Another categories of FRB progenitor systems invoke catastrophic events involving compact star mergers 
such as double neutron star, neutron star-black hole, double black hole mergers 
\citep[e.g.][]{Tot13, Pir12, zhang16b,zhang16a,Wa16, Li16}.
Star formation process is usually not relevant in such events.
So, unless the merger delay time scale is shorter than Myr, as expected in some prompt merger scenarios,
the scattering mechanism introduced in this work does not apply to these FRBs.

In contrast to the consideration of extensively distributed scattering medium in this work, 
the ad hoc thin screen scattering model applies when the scattering matter is concentrated in a local region. 
By assuming a uniform distribution of the density irregularities along the LOS through the scattering region, a thick scattering screen behaves similar to a thin 
screen, except that the depth passing through the extended scattering medium should be replaced by a much smaller screen thickness in the latter case. 
According to Eq. \eqref{eq: tscsimp},
extraordinary high density contrast is required to compensate for the dramatic decrease of $D$ and account for strong scattering. 
This localized density excess is too large to be produced and confined in diffuse IGM or ISM
\citep{Kat14},
but could possibly be associated with the FRB source and located in its immediate vicinity
\citep{Mas15}. 
For this reason, we relate the thin screen scattering scenario to the intrinsic pulse width and exclude it from our analysis on the scattering effect arising in more 
diffuse media.

The microscale instabilities are an important physical ingredient in many fundamental
processes such as heat conduction
\citep{Chan98}, 
dynamo growth of magnetic fields
\citep{Sch06}, 
and acceleration of cosmic rays
\citep{LB06} 
in the IGM. 
The evolution of instabilities are directly related to the magnetic field geometry and intensity at scales smaller than the particle mean free path. 
Multiple observational techniques have been utilized to measure the extragalactic large-scale ($> 1$ kpc) magnetic fields
\citep{Kro94, Car02, Gov04, Xu06}, 
but detailed information on small-scale magnetic field structures is still unaccessible due to the limited spatial resolution.
As exemplified in this work, the pulse durations of FRBs pose an upper bound on the 
amplitude of density and magnetic fluctuations, and a lower bound on their characteristic scale, which can be potentially
exploited as an observational approach of studying the properties of collisionless regime of the IGM turbulence.

The sheet-like structures of density in the viscosity damped regime of MHD turbulence are unlikely to dominate the strong scattering of radio waves 
as suggested in earlier studies
(e.g., \citealt{GS06}). 
In the presence of cascade of MHD turbulence, not only the rigidity of the folded magnetic field structure can easily break down, but also the local 
magnetic variation on the viscous scale fails to produce sufficient density fluctuations.

\acknowledgments
We thank the anonymous referee for helpful suggestions, and Jonathan Katz, Dick Manchester, and Divya Palaniswamy for useful comments. This work is partially supported by the National Basic Research Program (973 Program) of China under grant No. 2014CB845800.

\appendix

\section{Generalized formalism of temporal broadening}\label{app}

The derivation of the diffractive scattering formalism
presented in Section \ref{eq: secpw} for a power-law spectrum of electron density fluctuations can be further generalized. 
We first write the phase structure function as 
\begin{equation}\label{eq: sfchasc}
    D_\Phi \sim \pi r_e^2 \lambda^2 D (\delta n_e(r))^2 r.
\end{equation} 
At the diffractive scale $r_\text{diff}$, $D_\Phi = 1$ is satisfied and there is  
\begin{equation}\label{eq: screla}
  \pi  r_e^2 \lambda^2 D (\delta n_e(r_\text{diff}))^2 r_\text{diff} = 1. 
\end{equation}
Substituting $r_\text{diff}^{-1} $ from the above equation in Eq. \eqref{eq: tscpw} leads to a general form of the scattering timescale, 
\begin{equation}\label{eq: gentsc}
  \tau_{sc}  = \frac{D \lambda^2}{4\pi^2 c} r_\text{diff}^{-2} 
                  = \frac{D^2 r_e^2 \lambda^4}{4\pi c}  \frac{(\delta n_e(r_\text{diff}))^2}{ r_\text{diff}}.
\end{equation}

We first use Eq. \eqref{eq: gentsc} to reproduce the expressions of $\tau_{sc}$ corresponding to a spatial power spectrum of density fluctuations
derived in Section \ref{eq: secpw}. 
In the case of $r_\text{diff} < l_0$, the scattering effect is dominated by the 
inner scale $l_0$ of the density spectrum. 
From Eq. \eqref{eq: dsfgau} and \eqref{eq: sfchasc}, the diffractive scale is
\begin{equation}\label{eq: rdgo}
   r_\text{diff} = \frac{l_0}{\sqrt{D_\Phi (l_0)}} 
   =   [\pi r_e^2 \lambda^2 D (\delta n_e(l_0))^2 ]^{-\frac{1}{2}} l_0^\frac{1}{2},
\end{equation}
where the electron density perturbation at $l_0$ depends on the spectral shape, 
\begin{subnumcases}
    {(\delta n_e(l_0))^2 =\label{eq: denss}}
     (\delta n_e)^2 \Big(\frac{l_0}{L}\Big)^{\beta-3},   ~~ \beta > 3, \\
     (\delta n_e)^2, ~~~~~~~~~~~~~~~~~ \beta < 3.
\end{subnumcases}
Thus $r_\text{diff}$ has the form 
\begin{subnumcases}
    {r_\text{diff} =\label{eq: rdrec}}
    \Big( \pi r_e^2 \lambda^2 D (\delta n_e)^2 L^{3-\beta} l_0^{\beta-4}\Big)^{-\frac{1}{2}}, ~ \beta > 3, \\
    \Big(\pi r_e^2 \lambda^2 D (\delta n_e)^2  l_0^{3-\beta} l_0^{\beta-4}\Big)^{-\frac{1}{2}} ,~~ \beta < 3, 
\end{subnumcases}
It recovers Eq. \eqref{eq: rmina} in combination with Eq. \eqref{eq: cnsss} and \eqref{eq: simsm}. 
Using Eq. \eqref{eq: screla} together with Eq. \eqref{eq: rdgo},  we find 
\begin{equation}
  \frac{(\delta n_e(r_\text{diff}))^2}{ r_\text{diff}} = (\pi  r_e^2 \lambda^2 D )^{-1} r_\text{diff}^{-2} = \frac{(\delta n_e(l_0))^2}{l_0}.
\end{equation}
Inserting this into Eq. \eqref{eq: gentsc} and considering Eq. \eqref{eq: denss} yields 
\begin{subnumcases}
    {\tau_{sc} =}
     \frac{D^2 r_e^2 \lambda^4}{4\pi c} (\delta n_e)^2 \Big(\frac{l_0}{L}\Big)^{\beta-4} L^{-1}, ~~ \beta > 3, \\
     \frac{D^2 r_e^2 \lambda^4}{4\pi c} (\delta n_e)^2 l_0^{-1}, ~~~~~~~~~~~~~~~~~~ \beta < 3,
\end{subnumcases}
which after we incorporate the $(1+z_\text{q})$ factor and replace $D$ with $D_\text{eff}$ 
have the same expressions as $W$ in Eq. \eqref{eq: scste} and \eqref{eq: scsha}.

When $r_\text{diff}$ resides within the inertial range, $r_\text{diff} > l_0$, the density perturbation at $r_\text{diff}$ can be given according to the 
power-law scaling of the spectrum, 
\begin{subnumcases}
    {(\delta n_e(r_\text{diff}))^2 =}
     (\delta n_e)^2 \Big(\frac{r_\text{diff}}{L}\Big)^{\beta-3},   ~~~ \beta > 3, \\
     (\delta n_e)^2 \Big(\frac{r_\text{diff}}{l_0}\Big)^{\beta-3}, ~~~ \beta < 3.
\end{subnumcases}
It can be equivalently written as
\begin{equation}\label{eq: nesm}
   (\delta n_e(r_\text{diff}))^2 = \frac{\text{SM}}{D} r_\text{diff}^{\beta-3}. 
\end{equation}
Substituting this into Eq. \eqref{eq: screla} gives 
\begin{equation}\label{eq: rdsm}
    r_\text{diff} = \Big(\pi r_e^2 \lambda^2 \text{SM}\Big)^{\frac{1}{2-\beta}}, 
\end{equation}
which recovers Eq. \eqref{eq: rminb}.
From both Eq. \eqref{eq: nesm} and \eqref{eq: rdsm}, we can now get 
\begin{equation}
  \frac{(\delta n_e(r_\text{diff}))^2}{ r_\text{diff}} = \frac{\text{SM}}{D} r_\text{diff}^{\beta-4} 
  =  \Big(\pi r_e^2 \lambda^2 \Big)^\frac{4-\beta}{\beta-2} D^{-1} \text{SM}^\frac{2}{\beta-2}.
\end{equation}
Thus $\tau_{sc}$ from Eq. \eqref{eq: gentsc} in this case becomes 
\begin{equation}
\tau_{sc}  = \frac{D r_e^\frac{4}{\beta-2} \lambda^\frac{2\beta}{\beta-2}}{4\pi^\frac{2(\beta-3)}{\beta-2} c}    \text{SM}^\frac{2}{\beta-2}.
\end{equation}
We can further derive (Eq. \eqref{eq: cnsss} and \eqref{eq: simsm})
\begin{subnumcases}
{\tau_{sc}  =}
\frac{D^\frac{\beta}{\beta-2} r_e^\frac{4}{\beta-2} \lambda^\frac{2\beta}{\beta-2}}{4\pi^\frac{2(\beta-3)}{\beta-2} c}  (\delta n_e)^\frac{4}{\beta-2} L^\frac{2(3-\beta)}{\beta-2}, 
~~~~~~~ \beta > 3, \\
\frac{D^\frac{\beta}{\beta-2} r_e^\frac{4}{\beta-2} \lambda^\frac{2\beta}{\beta-2}}{4\pi^\frac{2(\beta-3)}{\beta-2} c}  (\delta n_e)^\frac{4}{\beta-2} l_0^\frac{2(3-\beta)}{\beta-2}, 
~~~~~~~~~\beta < 3.
\end{subnumcases}
After adding the $(1+z_\text{q})$ factor to the above expressions and using $D_\text{eff}$ instead of $D$, we obtain the same results in the observer's frame
as in Eq. \eqref{eq: scsteb} and \eqref{eq: scshab}.

When the density irregularities are characterized by a density perturbation $\delta n_e(d)$ and a length scale $d$, 
similar to the case of a density power spectrum with $r_\text{diff} < l_0$, the phase structure function can be simplified 
\citep{Sch68},
\begin{equation}
    D_\Phi \sim \pi r_e^2 \lambda^2 D (\delta n_e(d))^2 d.
\end{equation} 
Strong scattering occurs when $d$ exceeds $r_\text{diff}$, which is 
\begin{equation}\label{eq: sigsrs}
   r_\text{diff} = \frac{d}{\sqrt{D_\Phi }} 
   =   [\pi r_e^2 \lambda^2 D (\delta n_e(d))^2 ]^{-\frac{1}{2}} d^\frac{1}{2}. 
\end{equation}
The condition $d > r_\text{diff}$ (i.e. $\sqrt{D_\Phi} > 1$) sets a minimum $d$ when $\delta n_e(d)$ is provided, 
\begin{equation}\label{eq: mimd}
   d > [\pi r_e^2 \lambda^2 D (\delta n_e(d))^2 ]^{-1},
\end{equation}
or a minimum density perturbation at a given $d$, 
\begin{equation}\label{eq: mimde}
    (\delta n_e(d))^2 >  [\pi r_e^2 \lambda^2 D d ]^{-1}.
\end{equation}
From the relation Eq. \eqref{eq: screla} and the expression of $r_\text{diff}$ in Eq. \eqref{eq: sigsrs}, we get 
\begin{equation}
  \frac{(\delta n_e(r_\text{diff}))^2}{ r_\text{diff}} = (\pi  r_e^2 \lambda^2 D )^{-1} r_\text{diff}^{-2} = \frac{(\delta n_e(d))^2}{d}.
\end{equation}
So the general form of $ \tau_{sc}$ in Eq. \eqref{eq: gentsc} in this situation becomes 
\begin{equation}\label{eq: tscsimp}
  \tau_{sc}  = \frac{D^2 r_e^2 \lambda^4}{4\pi c}   \frac{(\delta n_e(d))^2}{d}.
\end{equation}

At the observer's wavelength $\lambda_0$, 
Eq. \eqref{eq: mimd}, \eqref{eq: mimde} become 
\begin{equation}\label{eq: mimdoapp}
   d > (1+z_\text{q})^2 [\pi r_e^2 \lambda_0^2 D_\text{eff} (\delta n_e(d))^2 ]^{-1},
\end{equation}
\begin{equation}\label{eq: mimdeoapp}
    (\delta n_e(d))^2 >  (1+z_\text{q})^2 [\pi r_e^2 \lambda_0^2 D_\text{eff} d ]^{-1},
\end{equation}
and the pulse scatter-broadening measurement from Eq. \eqref{eq: tscsimp} in the frame of the observer is 
\begin{equation}\label{eq: tscsimpoapp}
  W  =   \frac{D_\text{eff}^2 r_e^2 \lambda_0^4}{4\pi c (1+z_\text{q})^3}   \frac{(\delta n_e(d))^2}{d},
\end{equation}
which in terms of the DM caused by the scattering medium is 
\begin{equation}
  W  =   \frac{ r_e^2 \lambda_0^4}{4\pi c (1+z_\text{q})^3}   \Big(\frac{\delta n_e(d)}{n_e}\Big)^2 d^{-1} \text{DM}^2.
\end{equation}
Eq. \eqref{eq: mimdoapp}-\eqref{eq: tscsimpoapp} impose observational constraints on  
the density perturbation and its characteristic scale that the density fluctuation model under consideration must satisfy.

\bibliographystyle{apj.bst}
\bibliography{yan}

\end{document}